\documentclass[aps,prl,twocolumn,showpacs,
nolongbibliography,10pt]{revtex4-2}

\usepackage[colorlinks = true,linkcolor = red,citecolor = magenta]{hyperref}
\usepackage[sort&compress]{natbib}
\usepackage{scalerel}
\usepackage[normalem]{ulem}
\usepackage{xcolor}
\usepackage{bm}
\usepackage{amsmath}
\usepackage{amsfonts}
\usepackage{amssymb}
\usepackage{graphicx}
\usepackage{subfigure}
\usepackage{mathtools}
\usepackage{physics}
\usepackage{dsfont}
\usepackage{float}
\usepackage{lmodern}
\usepackage[many]{tcolorbox}
\usepackage{empheq}
\usepackage{cleveref}
\usepackage{textcomp}
\newcommand{\scs}{\scriptscriptstyle}

\begin{document}

\title{Quantum Noise Suppression in Non-Hermitian Resonators at Exceptional Point}
\author{Dmitrii N. Maksimov$^{1,2,3,*}$}
\author{Andrey A. Bogdanov$^{1,4,\dagger}$}
\affiliation{$^1$Qingdao Innovation and Development
Center, Harbin Engineering University, Qingdao 266000, Shandong, China}
\affiliation{$^3$Kirensky Institute of Physics, Federal Research Centre KSC SB RAS, 660036, Krasnoyarsk, Russia}
\email{mdn@tnp.krasn.ru}
\affiliation{$^2$IRC SQC, Siberian Federal University, 660041, Krasnoyarsk, Russia,}
\affiliation{$^4$School
of Physics and Engineering, ITMO University, 197101, St.~Petersburg, Russia}
\email{a.bogdanov@hrbeu.edu.cn}

\date{\today}

\begin{abstract}
We investigate the impact of quantum noise on non-Hermitian resonators at an exceptional point (EP). The system's irreversible Markovian dynamics is modeled using the Lindblad master equation, which accounts for the incoherent pump, radiative losses, and external monochromatic field. An exact analytic solution is derived in the form of the characteristic function of the Husimi distribution, enabling the calculation of all quantum mechanical observables associated with the bosonic degrees of freedom. Our analysis reveals that quantum noise strongly influences the system's response when the system exhibits $\mathcal{P}\mathcal{T}$-symmetry. Out of the $\mathcal{P}\mathcal{T}$-symmetric regime, however, the system demonstrates stability within a specific parametric domain, where the effects of quantum noise on the signal-to-noise ratio can be mitigated by increasing the external field. 
\end{abstract}
\maketitle


Recent advances in non-Hermitian quantum systems have opened new avenues for revolutionary applications in quantum sensing, quantum metrology, and amplification \cite{feng2017non, el2019dawn, Wiersig2020a, Wang2023}. Among the defining features of these systems is the phenomenon of {\it exceptional points} (EPs), where two or more eigenvalues and their associated eigenstates coalesce, resulting in profound changes in the system's dynamics~\cite{Miri2019Jan}. EPs not only are of fundamental interest, but also have significant potential for practical applications, particularly in high-precision measurements and quantum sensors. Systems operating near EPs exhibit an enhanced sensitivity to external perturbations, which can be harnessed to detect subtle changes in physical parameters, such as electromagnetic fields, temperature, or pressure \cite{Liu2016, Hodaei2017, xiao2019enhanced, McDonald2020, Chu2020, Luo2022, Kononchuk2022}.

EPs are often associated with the concept of $\mathcal{P}\mathcal{T}$-symmetry, where gain and loss are precisely balanced in a system. However, achieving an EP does not necessarily require strict compliance with this balance. The $\mathcal{P}\mathcal{T}$-symmetric contribution to the system's Hamiltonian can be separated, enabling the construction of EPs in systems where all eigenstates exhibit decaying behavior~\cite{Miri2019Jan}. This  allows for broader flexibility in the design of systems operating at or near EPs as the specific realization of gain-loss balance becomes less critical.
\begin{figure}[t]
\includegraphics[width=1\linewidth]{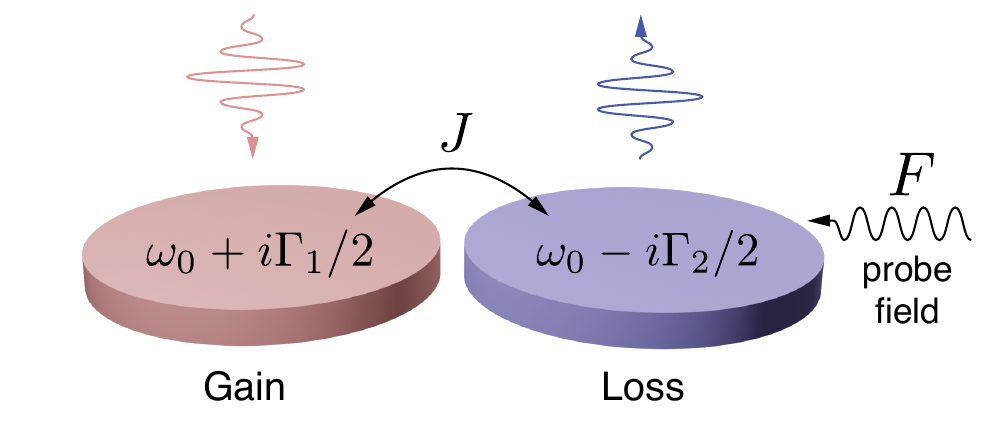}
\caption{Two coupled resonators with eigenfrequency $\omega_0$. The left resonator is pumped at rate $\Gamma_1$. The right resonator is open for radiation losses at rate $\Gamma_2$ and is subject to external monochromatic driving $F$ acting as a probe field.}
\label{fig1}
\end{figure}

The realization of enhanced quantum sensing capabilities is constrained by the influence of {\it quantum noise}, a key factor in open quantum systems~\cite{ Breu02, gardiner2014quantum} that is inherent to the irreversible dynamics under the evolution semigroup \cite{gisin1992quantum, Dale14}. 
 Although the response near EPs is characterized by enhanced sensitivity to external perturbations, the effect of quantum noise sets fundamental limits on measurement precision \cite{yoo2011quantum, Lau2018, langbein2018no, Chen2019, simonson2022nonuniversality, zhang2019quantum, Wang2020, Ding2023, Wang2024,Loughlin2024Jun} leading to a degraded {\it signal-to-noise ratio} (SNR) as discussed in \cite{langbein2018no}. The interplay
between sensitivity enhancement and noise amplification is a critical factor that determines the practical utility of EPs in sensing applications. The role of fluctuations and their impact on the performance of EP-based sensors remain topics of active investigation~\cite{Mortensen2018Oct,Wiersig2020a,zhang2019quantum,Kononchuk2022,xiao2019enhanced,Mao2024Apr}.


In this work we consider the system of two coupled single-mode quantum resonators one of which is subject to photonic gain while the other is interacting with an external monochromatic field. The non-Hermitian model in our analysis can be viewed as the Scully-Lamb model of laser \cite{arkhipov2019scully, arkhipov2020quantum} away from the gain saturation (lasing) threshold. Following \cite{arkhipov2019scully, Longhi2020, Roccati2021, Roccati2023, nakanishi2022pt,larson2023exceptional,Chen2021,Minganti2019, abo2024experimental, arkhipov2020quantum}, we analyze the problem by solving a Lindbladian master equation which is regarded as the most reliable tool since it respects the key properties of the system's density matrix such as non-negativity, Hermitisity, and normalization \cite{gardiner2014quantum}. It is worth mentioning that Lindbladian master equations do not only account for the genuine quantum noise but can also accommodate parametric noise as shown in \cite{Wiersig2020May}. Our goal is to derive an analytic solution for the model, enabling the calculation of arbitrary-order moments with respect to the creation and annihilation operators acting on the Fock space of the two resonators.

\nopagebreak

The system under consideration is sketched in Fig.~\ref{fig1} where we show two coupled single-mode resonators with equal eigenfrequencies $\omega_0$. One of the resonators is subject to bosonic gain whereas the other permits radiation losses to the outer space. The system is driven by an external monochromatic field with frequency $\omega$ applied to the lossy resonator. The dynamics is controlled by the master equation for the reduced density matrix $\widehat{\mathcal{R}}$ in the Lindblad form
\begin{equation}\label{master}
\frac{\partial \widehat{\mathcal{R}}}{\partial t}=
-i[\widehat{\mathcal{H}},\widehat{\mathcal{R}} ]+
\widehat{\mathcal{L}}_g(\widehat{\mathcal{R}})+
\widehat{\mathcal{L}}_d(\widehat{\mathcal{R}}).
\end{equation}
The Hamiltonian $\widehat{\mathcal{H}}$ is as follows
\begin{align}
& \widehat{\mathcal{H}}\!=\!\Delta\hat{a}_1^{\dagger}\hat{a}_1\!+\!
\Delta\hat{a}_2^{\dagger}\hat{a}_2\! 
+\!\frac{J}{2}\!\left(
\hat{a}_2^{\dagger}\hat{a}_1
\!+\!\hat{a}_1^{\dagger}\hat{a}_2
\right)\!+\!F(\hat{a}_2\!+\!\hat{a}_2^{\dagger}),
\end{align}
where $\hat{a}_{1,2}$ are the bosonic annihilation operators in the first and the second resonators, correspondingly,  $J$ is the tunneling matrix element between the resonators, $\Delta=\omega_0-\omega$ is the frequency detuning, and $F$ is the overlap integral between the eigenmode of the lossy resonator and the incident field. The external monochromatic driving plays the role of a probe field as shown in Fig.~\ref{fig1}. The gain $\widehat{\mathcal{L}}_g(\widehat{\mathcal{R}})$ and the drain $\widehat{\mathcal{L}}_d(\widehat{\mathcal{R}})$ bosonic Liouvillians are defined below
\begin{align}
& \widehat{\mathcal{L}}_g(\widehat{\mathcal{R}})=-\frac{\Gamma_1}{2}
\left(\hat{a}_1\hat{a}_1^{\dagger}\widehat{\mathcal{R}}
-2\hat{a}_1^{\dagger}\widehat{\mathcal{R}}\hat{a}_1+
\widehat{\mathcal{R}}\hat{a}_1\hat{a}_1^{\dagger}\right), \nonumber \\
& \widehat{\mathcal{L}}_d(\widehat{\mathcal{R}})=-\frac{\Gamma_2}{2}
\left(\hat{a}_2^{\dagger}\hat{a}_2\widehat{\mathcal{R}}
-2\hat{a}_2\widehat{\mathcal{R}}\hat{a}_2^{\dagger}+
\widehat{\mathcal{R}}\hat{a}_2^{\dagger}\hat{a}_2\right),
\end{align}
where $\Gamma_1$ is the pump rate in the first resonator and
$\Gamma_2$ is the decay rate in the second resonator.

In order to analytically solve Eq.~\eqref{master} for the observable quantities, we calculate the characteristic function of the Husimi quasi-probability distribution \cite{vogel1989quasiprobability, gardiner2014quantum}. The characteristic function is defined as
\begin{equation}\label{chi}
\chi(\alpha_1,\alpha_2,
\alpha_1^*,\alpha_2^*)=
\mathrm{tr}
\left(
e^{-\alpha_1^*\hat{a}_1-\alpha_2^*\hat{a}_2}
e^{\alpha_1\hat{a}_1^{\dagger}+
\alpha_2
\hat{a}_2^{\dagger}}
\widehat{{\mathcal{R}}}\right).
\end{equation}
This characteristic function contains the full information on the observable quantities since the expectation value of antinormally ordered operators can be calculated as
\begin{align}\label{moments}
& \langle \hat{a}_1^{n_1}
\hat{a}_2^{n_2}
(\hat{a}_1^{\dagger})^{m_1}
(\hat{a}_2^{\dagger})^{m_2}
\rangle
= \nonumber \\
& \left.\left(-\frac{\partial}{\partial \alpha_1^*} \right)^{n_1}
\left(
-\frac{\partial}{\partial \alpha_2^*}
\right)^{n_2}
\frac{\partial^{m_1}}{\partial \alpha_1^{m_1}}
\frac{\partial^{m_2}}{\partial \alpha_1^{m_2}}
\chi(\boldsymbol{\alpha})\right|_{\boldsymbol{\alpha}=0},
\end{align}
where ${\boldsymbol{\alpha}}^{\intercal}=(\alpha_1,\alpha_2,
\alpha_1^*,\alpha_2^*)$.
The Husimi distribution is a projection of the density matrix on the tensor product of the Glauber coherent states \cite{gardiner2014quantum} for each resonator. The Husimi distribution is related to its characteristic function via the Fourier transform
\begin{align}\label{Husimi}
& Q(a_1,a_2,
a_1^*,a_2^*)= \nonumber \\
& \frac{1}{\pi^4}\int\limits_{-\infty}^{\infty}
d {\bf \alpha}d{\bf \alpha}^*
\chi(\boldsymbol{\alpha})
e^{a_1\alpha_1^*+a_2\alpha_2^*
-a_1^*\alpha_1
-a_2^*\alpha_2},
\end{align}
where $a_1,a_2,
a_1^*,a_2^*$ are the parameters of the coherent states playing the role of complex conjugated canonical phase space variables \cite{gardiner2014quantum}.
The equation for the characteristic function is obtained
by applying Eq.~\eqref{chi} to the master equation. The result reads
\begin{align}\label{chi_eq}
&\frac{\partial \chi}{\partial t}=
-i\Delta
\left(
-\alpha_1\frac{\partial\chi}{\partial\alpha_1}
-\alpha_2\frac{\partial\chi}{\partial\alpha_2}
+\alpha_1^*\frac{\partial\chi}{\partial\alpha_1^*}
+\alpha_2^*\frac{\partial\chi}{\partial\alpha_2^*}
\right) \nonumber \\ &
-i\frac{J}{2}
\left(
-\alpha_2\frac{\partial\chi}{\partial\alpha_1}
-\alpha_1\frac{\partial\chi}{\partial\alpha_2}
+\alpha_2^*\frac{\partial\chi}{\partial\alpha_1^*}
+\alpha_1^*\frac{\partial\chi}{\partial\alpha_2^*}
\right) \nonumber \\
& +iF(\alpha_2+\alpha_2^*)\chi + \frac{\Gamma_1}{2}
\left(
\alpha_1\frac{\partial\chi}{\partial\alpha_1}+
\alpha_1^*\frac{\partial\chi}{\partial\alpha_1^*}
\right)  \nonumber \\
 &-\frac{\Gamma_2}{2}
\left(
\alpha_2\frac{\partial\chi}{\partial\alpha_2}+
\alpha_2^*\frac{\partial\chi}{\partial\alpha_2^*}
\right) - \frac{\Gamma_2}{2} |\alpha_2|^2\chi.
\end{align}
The Fokker-Planck equation for the Husimi function can be obtained
by Fourier transforming Eq.~\eqref{chi_eq} according to Eq.~\eqref{Husimi}. This, however, is not needed for our purpose. It is worth noting, though, that the last term in Eq.~\eqref{chi_eq} is quadratic with respect to the arguments of the characteristic function. Upon the Fourier transform this would lead to a second derivative in the Fokker-Planck equation. The second derivative is equivalent a noisy (diffusive) term in the corresponding Langevin-It\^o drift-diffusion equation \cite{vogel1989quasiprobability, gardiner2014quantum}. Thus, Eq.~\eqref{chi_eq} captures the effect of quantum noise. The formal stationary solution of Eq.~\eqref{chi_eq} reads
\begin{equation}\label{char}
\chi=\exp(
\langle q|\alpha\rangle-\langle\alpha| q\rangle+\langle\alpha|\widehat{M}|\alpha
\rangle),
\end{equation}
where $|\alpha\rangle=(\alpha_1,\alpha_2)^{\intercal}$,
\begin{align}
& \widehat{M}=
M_0
\left\{
\begin{array}{cc}
-J\Gamma_2 & -i\Gamma_1\Gamma_2 \\
i\Gamma_1\Gamma_2 & -J\Gamma_1-{M_0}^{-1}
\end{array}
\right\}, \nonumber \\ &\ M_0=\frac{J}{(\Gamma_2-\Gamma_1)(J^2-\Gamma_1\Gamma_2)},
\end{align}
and
\begin{align}
& |q\rangle=\frac{2F}{J^2-4\mathcal{E}_1\mathcal{E}_2}
\left\{
\begin{array}{c}
-J \\
2\mathcal{E}_1
\end{array}
\right\}, \nonumber \\
&\mathcal{E}_1=\Delta+i\frac{\Gamma_1}{2}, \ \
\mathcal{E}_2=\Delta-i\frac{\Gamma_2}{2}.
\end{align}
Note that Eq.~\eqref{char} complies with the normalization condition $\mathrm{tr}
(\widehat{{\mathcal{R}}})=1$.

\begin{figure}[t]
\includegraphics[width=0.4\textwidth,height=.35\textwidth,trim=2.2cm 7.3cm 2.5cm 7.4cm,clip]{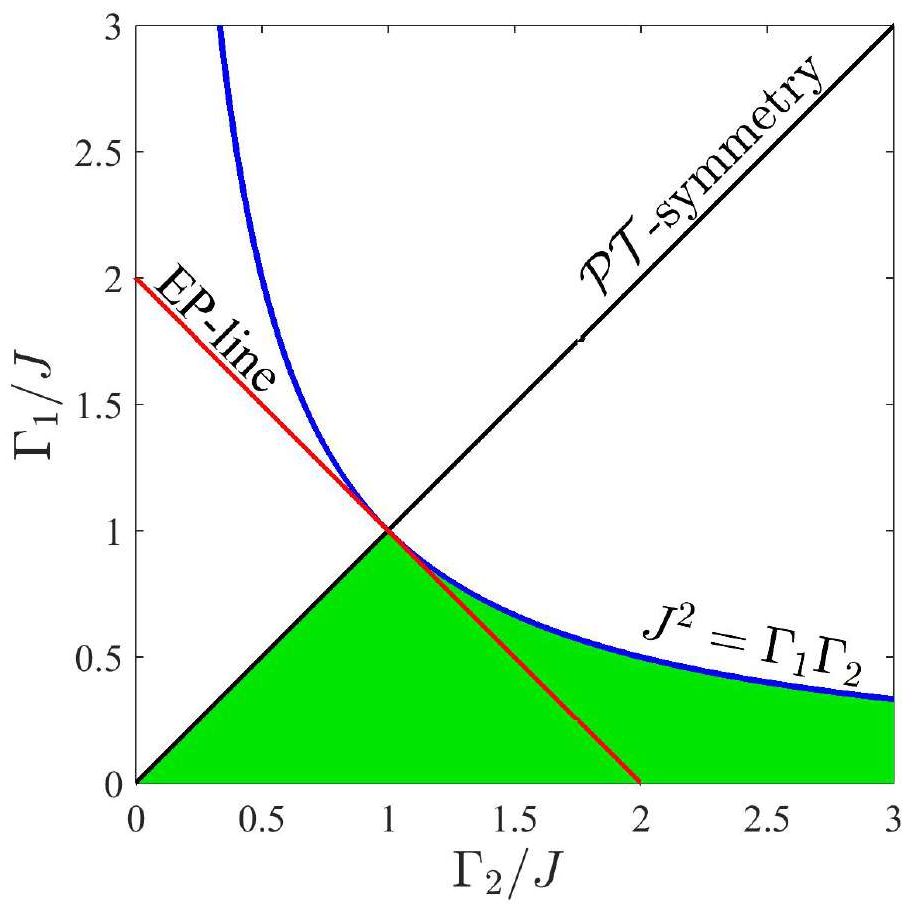}
\caption{Singularities in the parametric space of the system leading to instability according to Eq.~\eqref{stab}. EP-line is the line given by Eq.~\eqref{EPline}. The stability domain of the solution given by Eq.~\eqref{bara} is shaded in green.}
\label{fig2}
\end{figure}

Before discussing the SNR, let us analyze our findings in more detail. First, let us define a vector 
\begin{equation}\label{order}
|\bar{a}\rangle=\left[\mathrm{tr}(\hat{a}_1
\widehat{\mathcal{R}}), \ \mathrm{tr}(\hat{a}_2
\widehat{\mathcal{R}})\right]^\intercal.
\end{equation}
Since the $\mathrm{U}(1)$ symmetry is broken by the external monochromatic field $|\bar{a}\rangle$ is not equal to zero. It can be thought of as the order parameter and, thus, constitutes the coherent fraction of the system's response.
By direct application of Eq.~\eqref{order} to the master equation, Eq.~\eqref{master}, one find that $|\bar{a}\rangle$ satisfies the following equation
\begin{equation}\label{order_eq0}
i\frac{\partial |\bar{a}\rangle}{\partial t}= \widehat{H}_{\mathrm{eff}}|\bar{a}\rangle+
|F\rangle,
\end{equation}
where the effective non-Hermitian Hamiltonian \cite{rotter2009non, Dale14} $\widehat{H}_{\mathrm{eff}}$ and the external driving $|F\rangle$ are given by 
\begin{equation}\label{order_eq}
\widehat{H}_{\mathrm{eff}}=
\left\{
\begin{array}{cc}
\Delta+i\frac{\Gamma_1}{2} & \frac{J}{2} \\
\frac{J}{2} & \Delta-i\frac{\Gamma_2}{2}
\end{array}
\right\}, \ \ |F\rangle=
\left\{
\begin{array}{c}
     0 \\
     F 
\end{array}
\right\}.
\end{equation}
The stationary solution of Eq.~\eqref{order_eq} is
\begin{equation}\label{bara}
|\bar{a}\rangle=|q\rangle,
\end{equation}
which is consistent with Eqs.~\eqref{moments} and \eqref{char}.
The condition for the EP of $\widehat{H}_{\mathrm{eff}}$ can be found as
\begin{equation}\label{EPline}
\Gamma_1+\Gamma_2=2J.
\end{equation}
Meanwhile, the stability analysis shows that the time-stationary solution $|\bar{a}\rangle=|q\rangle$ is stable if
\begin{equation}\label{stab}
\Gamma_2 > \Gamma_1, \ \mathrm{and} \ J^2>\Gamma_1\Gamma_2.
\end{equation}
The stability analysis is summarized in Fig.~\ref{fig2}, where we demonstrate the stability domain along with the line of EPs.
The intensities of the field in the resonators can not be calculated from $|\bar{a}\rangle$ alone since, besides the coherent fraction, the system also exhibits incoherent response due to quantum noise. To derive the intensities we use  
the single-particle density matrix (covariance matrix) defined as
\begin{equation}
\hat{\rho}=\sum_{n,n'}\mathrm{tr}(\hat{a}_{n'}^{\dagger}\hat{a}_n\widehat{\mathcal{R}})|n\rangle
\langle n'|,
\end{equation}
where $|n\rangle, \ n=1,2$ are the single photon states of the resonators.
Taking into account the antinormal ordering of operators Eqs.~\eqref{moments} and \eqref{char}
yield the stationary solution
\begin{equation}\label{rho}
\hat{\rho}=| q\rangle\langle q|-\widehat{M}-\widehat{{\mathbb{I}}}.
\end{equation}
At the same time, applying the definition of $\hat{\rho}$ to the master equation \eqref{master}, one finds that $\hat{\rho}$ satisfies
\begin{align}\label{master_rho}
\frac{\partial \hat{\rho}}{\partial t}\!=\!
-i[\widehat{H}_{\mathrm{eff}}\hat{\rho}\!-\!
\hat{\rho}\widehat{H}^{\dagger}_{{\mathrm{eff}}}]
\!+\!i\left(|q\rangle\!\langle{ F}|\!-\!|F\rangle\!\langle q|\right)\!+\!\Gamma_1\!|1\rangle\!\langle 1|.
\end{align}
One can see after some algebra that Eq.~\eqref{rho} is the stationary solution of Eq.~\eqref{master_rho}. The field intensities are given by the diagonal elements of $\hat{\rho}$, i.e. $\langle I_n \rangle=\langle \hat{a}^{\dagger}_n\hat{a}_n \rangle, \ \ n=1,2$.
The intensity in the lossy resonator can be found as
\begin{equation}\label{mean}
\langle I_2\rangle=|q_2|^2+J\Gamma_1M_0.
\end{equation}
At the same time for the pumped resonator we have
\begin{equation}
\langle I_1\rangle=|q_1|^2+J\Gamma_2M_0-1.
\end{equation}
One can check that the intensities are always positive if the stability condition Eq.~\eqref{stab} is satisfied. Let us use
$\bar{a}_n$ as the signal in the definition of the SNR, then we have
\begin{equation}
\mathrm{SNR}^{\scs{(1)}}_n=\frac{q_n}{\sqrt{\langle I_n \rangle-|q_n|^2}}.
\end{equation}
One can see from the above equations that the the amplitude of the coherent fraction of the response is strongly effected by noise at the boundaries of the stability domain, Eq.~\eqref{stab}, where $M_0\rightarrow\infty$ and the coherent fraction is masked by incoherent radiation
\begin{equation}
\mathrm{SNR}^{\scs{(1)}}_n=0, \ \mathrm{if} \ M_0\rightarrow\infty.
\end{equation}
Particularly, the above equation holds true in the $\cal{P}\cal{T}$-symmetric case $\Gamma_1=\Gamma_2$. On the other hand away from the stability domain boundaries the SNR can be enhanced by increase of the external field
\begin{equation}
\mathrm{SNR}^{\scs{(1)}}_n\propto F, \ \mathrm{if} \ F\rightarrow\infty.
\end{equation}

\begin{figure}[t]
\includegraphics[width=1\linewidth]{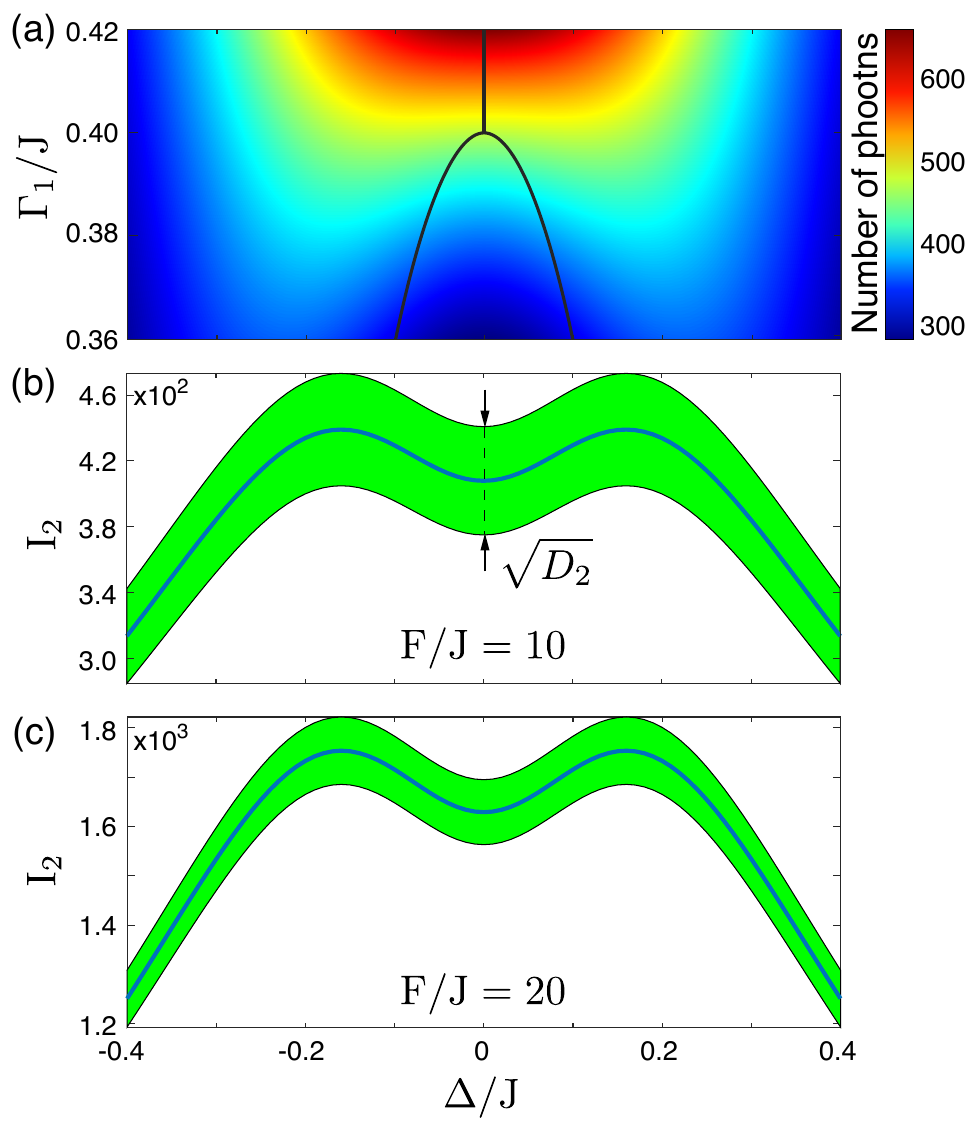}
\caption{Photon number (intensity) $I_2$ in the lossy resonator, $\Gamma_2/J=1.6$. (a) The intensity $I_2$ versus the detuning $\Delta$ and the pump rate $\Gamma_1$ for $F/J=10$. The black line shows the pitchfork bifurcation of the real parts of the eigenvalues of $\hat{H}_{\rm{eff}}$ in the EP.  (b, c) The intensity $I_2$ against the detuning $\Delta$ for two different values of the driving $F$ specified in the insets, $\Gamma_1/J=0.386$. The shaded areas show the mean deviation due to quantum noise.}
\label{fig3}
\end{figure}

The intensity dispersion at the $n$-th resonator is calculated as
\begin{equation}
D_n=\langle \hat{a}^{\dagger}_n\hat{a}_n\hat{a}^{\dagger}_n\hat{a}_n\rangle-\langle \hat{a}^{\dagger}_n\hat{a}_n\rangle^2, \ \ n=1,2
\end{equation}
After applying Eq.~\eqref{moments} in Eq.~\eqref{char} one finds
\begin{align}\label{dispersion}
& D_1=|q_1|^2(2J\Gamma_2M_0-1)+J\Gamma_2M_0(J\Gamma_2M_0-1), \nonumber \\
& D_2=|q_2|^2(2J\Gamma_1M_0+1)+J\Gamma_1M_0(J\Gamma_1M_0+1).
\end{align}
By defining the $\mathrm{SNR}^{\scs(2)}$ as
\begin{equation}
\mathrm{SNR}_n^{\scs(2)}=\frac{\langle I_n \rangle}{\sqrt{D_n}}
\end{equation}
and using Eqs.~\eqref{mean},~and~\eqref{dispersion} one finds again that the SNR can be enhanced by increasing 
the amplitude of the applied field, i.e.
\begin{equation}
\mathrm{SNR}_n^{\scs(2)}\propto F, \ \mathrm{if} \ F \rightarrow\infty,
\end{equation}
unless the system is at the boundaries of the stability domain, see Eq.~\eqref{stab}, where $M_0\rightarrow\infty$ and we have
\begin{equation}
\mathrm{SNR}_n^{\scs(2)}=1, \mathrm{if} \ M_0 \rightarrow\infty .
\end{equation}
Note that the intensity is not so strongly affected by noise as the order parameter. Our findings in regard to the system's response and the SNR are summarized in Fig.~\ref{fig3}.

In the EP, Eq.~\eqref{order_eq0} has the general solution
\begin{equation}\label{EP_solved}
|\bar{a}(t)\rangle\!=\!|q\rangle\!+\!\left(\!C_0|b\rangle \!+\!\frac{\sqrt{8}iC_1}{\Gamma_2\!+\!\Gamma_1}|2\rangle\!\right)e^{-i\lambda t}\!+\!C_1te^{-i\lambda t}|b\rangle,
\end{equation}
where $C_0,~C_1$ are complex numbers defined by the initial condition, while $\lambda$ and $|b\rangle$ given below
\begin{equation}
\lambda=\Delta-i\frac{\Gamma_2-\Gamma_1}{4}, \ \ |b\rangle=
\frac{1}{\sqrt{2}}
\left(
\begin{array}{c}
1 \\
-i
\end{array}
\right)
\end{equation}
solve the eigenvalue problem in the EP,
$ \widehat{H}_{\mathrm{eff}}|b\rangle=\lambda|b\rangle
$.
Next, by applying the definition of $\widehat{H}_{\mathrm{eff}}$ from Eq.~\eqref{order_eq} one can rewrite Eq.~\eqref{chi_eq} in a more convenient form
\begin{align}\label{chi_eq2}
& \frac{\partial \chi}{\partial t}=-i\left(\langle \alpha|\widehat{H}_{\mathrm{eff}}|\partial_{\alpha} \chi\rangle-
\langle \partial_{\alpha} \chi|\widehat{H}_{\mathrm{eff}}^{\dagger}|\alpha\rangle\right) \nonumber \\
&+i(\langle F|\alpha\rangle+\langle \alpha|F\rangle)\chi-
\frac{\Gamma_2}{2}|\langle\alpha|2\rangle|^2\chi,
\end{align}
where
\begin{equation}
\langle \partial_{\alpha} \chi|=
\left(\frac{\partial\chi}{\partial \alpha_1},\ \frac{\partial\chi}{\partial \alpha_2} \right).
\end{equation}
One can easily check that in the EP Eq.~\eqref{chi_eq2} has the following solution
\begin{align}\label{char2}
&\chi(t)= \int dC_0dC_1f(C_0,C_1) \nonumber \\
&\times\!e^{\langle \bar{a}(C_0,C_1,t)|\alpha\rangle-\langle\alpha|\bar{a}(C_0,C_1,t)\rangle+\langle\alpha|\widehat{M}|\alpha\rangle},
\end{align}
where
\begin{equation}
\int dC_0dC_1f(C_0,C_1)=1.
\end{equation}
The solution Eq.~\eqref{char2} satisfies boundary conditions
$\chi=0$ at $|\alpha\rangle\rightarrow\infty$ which guarantees convergence of the Fourier transform Eq.~\eqref{Husimi}. Equation~\eqref{char2} contains an arbitrary function of
tho complex parameters, $C_0$ and $C_1$. Since the phase space is four dimensional the function can be found from 
initial conditions for the Husimi distribution. Technically, though, one can simply set initial conditions for the momenta of the distributions (observables) and find $f(C_0,C_1)$ from Eq.~\eqref{moments}. 
Equation~\eqref{char2} describes the transient response in the EP. In the limit $t\rightarrow\infty$ Eq.~\eqref{char2} is identical to the stationary solution Eq.~\eqref{char} because all oscillatory terms in Eq.~\eqref{EP_solved} decay in time. The same conclusion holds everywhere in the stability domain \cite{comment}.

In summary -- the exact analytic solution away from the lasing threshold is found in the form of the characteristic function of the Husimi distribution, see Eq.~\eqref{char}. The characteristic function makes it possible to calculate all the quantum observables dependent on the field in the resonators. It is shown that the intensity response is strongly affected by quantum noise when the system is $\cal{P}\cal{T}$-symmetric in which case the SNR for the intensity limits to unity due to instability that manifests itself in increase of the incoherent fraction. Away from the $\cal{P}\cal{T}$-symmetry the solution is stable in the parametric domain specified by Eq.~\eqref{stab} where the SNR can be enhanced by increasing the external field. Of course, this approach is naturally limited by gain saturation and onset of lasing which is not accounted for in our solution. In regard to EPs, it is found that the general conclusion holds everywhere in the stability domain Eq.~\eqref{stab} whether an EP is present or not. The transient solution for the characteristic function in the EP is derived to ensure rigor, as shown in Eq.~\eqref{char2}. The time-dependent characteristic function is shown to limit to the stable stationary solution for any initial condition.

DNM acknowledges financial support from state contract No FWES-2024-0003 of Kirensky Institute of Physics.
\bibliography{Open_systems}

\begin{thebibliography}{42}
\bibitem [{com()}]{comment}%
  \BibitemOpen
  \href@noop {} {}\bibinfo {note} {Equation~\eqref{char2} can be generalized
  for any value of parameters allowing for a stable solution. To do this one
  has to solve the evolution equation for the order parameter, analogous to
  Eq.~\eqref{EP_solved}. This can be easily done away from the EP, where
  $\widehat{H}_{\mathrm{eff}}$ has two linearly independent eigenvectors, and
  the propagator can be written in the spectral representation.}\BibitemShut
  {Stop}%
\end{thebibliography}
\end{document}